\documentclass{revtex4}
\usepackage{graphicx}
\usepackage{fancyhdr}
\usepackage{amsmath}
\usepackage{wrapfig}
\usepackage{color}

\begin{document}

\title{Higgs sectors with exotic scalar fields\footnote{This talk is based on Refs.~\cite{kky,AKKY_full,AKKY,KY,AKY}, and it is presented as a title of ``Higgs Triplet Models".}}

%

\author{Kei Yagyu}
\affiliation{Department of Physics, National Central University, Chungli 32001, Taiwan}

\begin{abstract}

The discovery of the Higgs boson like particle with the mass of around 126 GeV has given us 
a great clue to know what is the true Higgs sector. 
New physics models at the TeV scale often introduce Higgs sectors extended from the minimal form, 
so that the determination of the Higgs sector can be a probe of new physics models. 
In this talk, we focus on the Higgs sector with exotic representation fields whose 
isospin is larger than 1/2. We first discuss the general features of exotic Higgs sectors, and then we consider several 
concrete models to clarify them. The phenomenology of the Higgs triplet model is discussed as the simple but important example. 

\end{abstract}

\maketitle

\thispagestyle{fancy}


\section{Introduction}

In July 2012, both the ATLAS and the CMS Collaborations have reported 
the discovery of a new boson with a mass of about 126 GeV~\cite{Higgs_LHC}. 
This particle has been observed in the $h\to \gamma\gamma$ and $h\to ZZ^*\to 4\ell$ modes, 
and those event numbers are compatible with that in the standard model (SM). 
Thus, the new particle is most likely the SM-like Higgs boson. 
However, this does not necessarily mean that the Higgs sector in the SM is correct, because 
the SM-like Higgs boson can also appear in the Higgs sector extended from the SM one. 

On the other hand, it has been known that there are phenomena which cannot be explained in the SM such as neutrino oscillations, 
the existence of dark matter and baryon asymmetry of the Universe. 
These phenomena are expected to be explained in models beyond the SM where extended Higgs sectors are often introduced, and their
properties strongly depend on the new physics scenario.  
Therefore, determining the structure of the Higgs sector is paramountly important to know 
what kind of the new physics models exists. 

The two Higgs doublet model (THDM) is one of the most popular and extensively analyzed extended Higgs sector. 
There are several scenarios of the THDM depending on the type of a discrete $Z_2$ symmetry~\cite{GW} which 
is basically introduced to avoid the tree level flavor changing neutral current (FCNC); $i.e.,$ the cases where 
the $Z_2$ symmetry is unbroken, softly-broken and/or explicitly-broken. 
The first case is known to be the inert doublet model~\cite{IDM} in which the lightest $Z_2$-odd scalar boson can be dark matter. 
In the second class of the THDM, there are four independent types of Yukawa interactions,
depending on the charge assignments of the $Z_2$ symmetry~\cite{4type}. 
The collider phenomenology can be drastically different among the four types of the Yukawa interactions~\cite{typeX}. 
The third scenario where the $Z_2$ symmetry is explicitly broken is so-called the type-III THDM. 
In this case, a particular Yukawa texture is introduced, forcing the non-diagonal
Yukawa couplings to be proportional to the geometric mean of the two fermion masses~\cite{typeIII}. 
As the other way to forbid the tree level FCNC, one can assume that one of the two Yukawa matrices 
is proportional to the other one, which is called as the aligned THDM~\cite{pich}. 
After the discovery of the SM-like Higgs boson, many literatures appeared in which the Higgs boson search data are explained based on 
the various scenarios in the THDM~\cite{recent_THDM}. 

Apart from the THDM, Higgs sectors which contains scalar fields with the isospin larger than 1/2 are also important to be studied.  
We here call such a Higgs field as an $exotic~Higgs~field$, and also such a Higgs sector as an $exotic~Higgs~sector$.  
The Higgs triplet model (HTM) is one of the simplest exotic Higgs sectors in which tiny neutrino masses are 
generated by the type-II seesaw mechanism~\cite{typeII}. 
In the HTM, the vacuum expectation value (VEV) of the triplet field deviates the electroweak rho parameter from unity at the tree level. 
The experimental value of the rho parameter is close to unity, so that the triplet VEV is severely constrained. 

There have been several models proposed with triplet Higgs fields. 
For instance, in the Georgi-Machacek (GM) model~\cite{GM,GM_pheno,Aoki_Kanemura},  
a real triplet Higgs field is added to the HTM in order to 
keep the electroweak rho parameter to be unity at the tree level by taking an alignment of two triplet VEV's.  
In the model proposed in Ref.~\cite{Kanemura_Sugiyama}, the lepton number violating mass in the HTM 
is induced at the one-loop level. 
In addition, the supersymmetric extention of the HTM has also been discussed in Ref.~\cite{typeII_susy}. 

As another example for exotic Higgs sectors, the Higgs quadruplet field is introduced in models proposed in Refs.~\cite{Quadruplet1,Quadruplet2},  
in which neutrino masses are generated through higher dimensional operators~\cite{Quadruplet1} and 
via the one-loop level~\cite{Quadruplet2}. 
The Higgs sector with the scalar quintuplet 
can be used to be built a two-loop radiative seesaw model~\cite{Quintuplet}. 
Recently, the model with the septet Higgs field has also been discussed in Refs.~\cite{Hisano_Tsumura,kky}.  

In this talk, we focus on exotic Higgs sectors. 
To know the $exoticness$ of the Higgs sector is important not only to probe exotic Higgs sectors but also to know the usual 
Higgs sector such as multi-doublet models. 
We first discuss the general features of the exotic Higgs sectors, and then we consider several concrete models. 

\section{Several features of exotic Higgs sectors}

There are several striking features in exotic Higgs sectors which can be listed as as follows: 
\begin{enumerate}
\item The electroweak rho parameter can deviate from unity at the tree level. 
\item The $H^\pm W^\mp Z$ vertex ($H^\pm$ are physical singly-charged Higgs bosons) appears at the tree level.  
The $H^\pm W^\mp \gamma$ vertex does not appear in any Higgs sectors at the tree level because of the $U(1)_{\text{em}}$ symmetry. 
\item The $hVV$ coupling ($h$ is the SM-like Higgs boson and $V=W$ or $Z$) can be larger than the SM value. 
\item An extra global $U(1)$ symmetry can accidentally exist in the Higgs potential. 
\end{enumerate}
Let us first define the general Higgs sector which contains $N$ Higgs multiplets 
$\Phi_i$ ($i=1,\dots , N$) with the isospin $T_i$ and the hypercharge $Y_i$ to clarify the properties 1, 2 and 3 listed in the above. 
The sum $T_i+Y_i$ corresponds to the electromagnetic charge $Q_i$, and it should be an integer number. 
We assume CP conservation of the Higgs sector for simplicity. 
The Higgs multiplet $\Phi_i$ can be expressed as 
\begin{align}
&\Phi_i = \left[\Phi_i^{Q=Y_i+T_i},\dots,\Phi_i^+, \Phi_i^0,\Phi_i^-,\dots \Phi_i^{Q=Y_i-T_i}\right]^T, \notag\\
&\text{with}~\Phi_i^0=\frac{1}{\sqrt{2c_i}}(h_i^0 +v_i+iz_i^0)~\text{for $Y_i\neq 0$ field},\quad\Phi_i^0=\frac{1}{\sqrt{2c_i}}(h_i^0 +v_i)~\text{for $Y_i= 0$ field},
\label{Phi}
\end{align}
where $v_i$ is the VEV of the Higgs multiplet, and $c_i=1~(1/2)$ for a complex (real) Higgs field. 
We note that in the Higgs field with $Y\neq 0$, 
the charge conjugation of the singly-charged component field $\Phi_i^+$ does not correspond to $\Phi_i^-$. 
The kinetic term in the general Higgs sector is given by 
\begin{align}
\mathcal{L}_{\text{kin}}=\sum_ic_i|D_i^\mu\Phi_i|^2,~\text{with}~D_i^\mu&=\partial^\mu-ig(T_i^+W^{+\mu} +T_i^-W^{-\mu})-i\frac{g}{c_W}(T_i^3-s^2_W Q_i)Z_\mu-ieQ_iA^\mu, \label{cov}
\end{align} 
where $T_i^\pm$ is the $SU(2)$ laddar operator, and $T_i^3$ is the third component of the isospin operator.
The weak mixing angle $\theta_W$ is 
introduced via $c_W=\cos\theta_W$ and $s_W=\sin\theta_W$.

\begin{center}
{\bf 1. The electroweak rho parameter}
\end{center}

From Eq.~(\ref{cov}), the $W$ and $Z$ boson masses are calculated as
\begin{align}
m_W^2=\frac{g^2}{4}v^2,\quad m_Z^2=\frac{g^2}{c^2_W}\sum_iv_i^2Y_i^2,~\text{with}~
v^2= 2\sum_i [T_i(T_i+1)-Y_i^2]v_i^2=(246~\text{GeV})^2. \label{vev}
\end{align}
The electroweak rho parameter can then be calculated at the tree level as~\cite{HHG}
\begin{align}
\rho_{\text{tree}}&=\frac{m_W^2}{m_Z^2c^2_W}
=\frac{\sum_i v_i^2[T_i(T_i+1)-Y_i^2]}{2\sum_i v_i^2Y_i^2}.\label{rho}
\end{align}
From this equation, we can find that the combination of $(T_i,Y_i)$ satisfied with $\rho_{\text{tree}}=1$ is 
\begin{align}
T_i=\frac{1}{2}\left(-1+\sqrt{1+12Y_i^2}\right). \label{rho1}
\end{align}
According to the above expression, the possible combinations are ($T_i,Y_i$)=(0,0), (1/2,1/2), (3,2), (25/2,15/2), $\cdots$. 
Thus, usually $\rho_{\text{tree}}$ deviates from unity in exotic Higgs sectors except for few special cases as written in just above. 
The magnitude of the deviation in $\rho_{\text{tree}}$ from unity depends on the value of the VEV of the exotic Higgs field. 
The experimental value of the rho parameter is give as $\rho_{\text{exp}}=1.0004^{+0.0003}_{-0.0004}$~\cite{PDG} so that 
the VEV of the exotic field is severely constrained by the data. 
For instance, in the Higgs sector with the complex (real) triplet field in addition to the usual doublet field, 
the upper bound for the triplet VEV is about 3.5 GeV (3.8 GeV) with the 95\% confidence level.  
There is the other way to satisfy $\rho_{\text{tree}}=1$ even in the Higgs sector contained 
Higgs multiplets without the relation given in Eq.~(\ref{rho1}).  
Namely, by taking some VEV's of exotic Higgs fields so as to keep the custodial symmetry, the rho parameter is kept to be unity 
at the tree level. 
The representative example is the GM model where the real and the complex triplet fields are added to the minimal Higgs sector, 
and VEV's of the triplet fields are taken to be aligned to keep the custodial symmetry. 
\begin{center}
{\bf 2. The $H^\pm W^\mp Z$ vertex}
\end{center}
In order to discriminate exotic Higgs sectors, we need to measure the other observables which are sensitive to the structure 
of the Higgs sector. 
A common feature in the extended Higgs sectors 
is the appearance of physical singly-charged Higgs bosons $H^\pm$.  
Hence, we may be able to discriminate each Higgs sector through 
the physics of charged Higgs bosons. 
Among the various observables related to $H^\pm$, 
the $H^\pm W^\mp Z$ vertex is useful to test exotic Higgs sectors. 
The magnitude of the $H^\pm W^\mp Z$ vertex can be parameterized by introducing $\xi_\alpha$ 
in the effective Lagrangian $\mathcal{L}=igm_W \xi_\alpha H_\alpha^+ W^- Z+\textrm{h.c.}$, where 
$H_\alpha^\pm$ is the $\alpha$-th physical singly-charged Higgs bosons.  
The $\xi_\alpha$ parameter is calculated at the tree level as~\cite{Grifols,HHG} as
\begin{align}
|F|^2 \equiv  \sum_{\alpha } |\xi_\alpha|^2 
&=\frac{2g^2}{c^2_Wm_W^2}
\Big\{\sum_i[T_i(T_i+1)-Y_i^2]v_i^2Y_i^2\Big\}-\frac{1}
{c^2_W\rho_{\textrm{tree}}^2}, 
\end{align}
where $\rho_{\rm tree}$ is given in Eq.~(\ref{rho}). 
A non-zero value of $|F|^2$ appears at the tree level 
only when $H_\alpha^\pm$ comes from an exotic Higgs field. 
In multi-doublet models, this vertex is induced at the one loop level, so that 
the magnitude of the vertex tends to be smaller than that in exotic Higgs sectors~\cite{HWZ_doublet}. 
Therefore, a precise measurement of the $H^\pm W^\mp Z$ vertex can be used to discriminate exotic Higgs sectors 
with $\rho_{\text{tree}}=1$ such as the GM model. 
The feasibility of measuring this vertex at collider experiments has been discussed in several papers~\cite{HWZ-LEP,HWZ-Tevatron,HWZ-LHC,epem,HWZ-ILC}. 
In Ref.~\cite{HWZ-LHC}, the single $H^\pm$ production from the $W^\pm Z$ fusion and 
the $W^\pm Z$ decay of $H^\pm$ have been analysed at the LHC. 
In the case with the mass of $H^\pm$ to be 200 GeV and $\mathcal{B}(H^\pm \to W^\pm Z)=100\%$, 
$|F|^2\gtrsim 0.036$ is required  to reach the signal significance to be larger than 2 at the collision energy and the integrated 
luminosity to be 14 TeV and 600 fb$^{-1}$, respectively. 
At the International Linear Collider (ILC), 
the $H^\pm W^\mp Z$ vertex can be measured via the $e^+e^-\to Z^*\to H^\pm W^\mp$ process~\cite{epem,HWZ-ILC}. 
In Ref.~\cite{HWZ-ILC}, this process has been studied in the lepton specific $H^\pm$ scenario; 
$i.e.,$ $\mathcal{B}(H^\pm\to \ell^\pm\nu)=100\%$. When the mass of $H^\pm$ is 150 GeV, 
$|F|^2 > 0.001$ is required to reach the signal significance to be larger than 2 with collision energy to be 300 GeV and 
integrated luminosity to be 1 ab$^{-1}$.

\begin{center}
{\bf 3. The $hVV$ vertex}
\end{center}

So far, we have not observed any new particles other than the ``Higgs boson" with the mass of 126 GeV, so that 
focusing on the Higgs boson couplings is quite important. 
First, we define the SM-like Higgs boson $h$ whose mass is taken to be 126 GeV in extended Higgs sectors by 
\begin{align}
h = (R^T)_{hi}h_i^0, 
\end{align}
where $R_{ih}$ is the element of the orthogonal matrix connecting $h_i^0$ given in Eq.~(\ref{Phi}) and the mass eigenstates for the CP-even 
scalar bosons. 
%
The $hVV$ couplings are then calculated by 
\begin{align}
g_{hVV} = g_{hVV}^{\text{SM}}\times \sum_i c_{hVV}^i
= g_{hVV}^{\text{SM}}c_{hVV}
,\quad \text{with }V=W,~Z, \label{g_hVV}
\end{align}
where $g_{hVV}^{\text{SM}}$ is the $hVV$ coupling in the SM, and  
the factor $c_{hVV}^i$ is expressed by
\begin{align}
c_{hWW}^i =  \frac{2v_i}{v}[T_i(T_i+1)-Y_i^2]R_{ih},\quad 
c_{hZZ}^i =  \frac{2Y_i^2 v_i R_{ih}}{\sqrt{\sum_j Y_j^2 v_j^2}}. \label{c_hVV}
\end{align}

In the general Higgs sector, the charged (neutral) 
Nambu-Goldstone (NG) bosons $G^\pm$ ($G^0$) can be separated from physical charged Higgs bosons (CP-odd Higgs bosons) 
by using the elements of the orthogonal matrices; 
\begin{align}
\Phi_i^\pm = R_{iG^+}G^\pm,\quad z_i^0 = R_{iG^0}G^0, \text{ with }\sum_i R_{iG^+}^2=\sum_iR_{iG^0}^2=1.\label{RG}
\end{align}
If the hypercharge of $\Phi_i$ is $Y_i\neq 0$ and $Y_i\neq T_i$, then $(\Phi_i^-)^*$ should also be included in the vector $\Phi_i^+$, and 
the corresponding matrix element $R_{iG^+}$ should be added. 

From the NG theorem, $R_{iG^+}$ and $R_{iG^0}$ satisfy the following relations; 
\begin{align}
\frac{g}{2}\sum_i \sqrt{c_i}C_i v_i R_{iG^+}=m_W,~ \frac{g}{c_W}\sum_i Y_i v_i R_{iG^0}=m_Z,~\text{with}~
C_i =\sqrt{T_i(T_i+1)-Y_i^2+Y_i}. 
\end{align}
%

We note that the $c_{hVV}$ factor in multi-doublet models is expressed by using Eqs.~(\ref{g_hVV}) and (\ref{c_hVV}) as  
\begin{align}
c_{hVV}=\sum_{i}\frac{v_iR_{ih}}{v}.
\end{align}
We can see that the factor is smaller than 1 because of the sum rule $\sum_i v_i^2=v^2$ (see Eq.~(\ref{vev})) in multi-doublet models. 
However, this feature $c_{hVV}\leq 1$ does notnecessarily hold in exotic Higgs sectors. 
We will see a few models with $c_{hVV}\geq 1$ in the next section. 

\begin{center}
{\bf 4. Global $U(1)$ symmetry}
\end{center}

The last feature of exotic Higgs sectors listed in the begging of this section is regarded with an extra global $U(1)$ symmetry. 
Let us consider the Higgs sector composed from one exotic Higgs field $X$ in addition to the Higgs doublet filed. 
If $X$ has quantum numbers of $T > 3/2$ and $Y\neq 0$ or $T= 3/2$ and $Y\neq \pm 3/2$, the Higgs sector has an global $U(1)$ symmetry 
associated with the phase rotation of $X$~\cite{logan}. 
If this symmetry is spontaneously broken down due to a non-zero VEV of $X$, then a massless NG boson appears in addition to 
the usual three NG bosons $G^\pm$ and $G^0$.  
A model with such an additional NG boson is phenomenologically unacceptable because the NG boson can couple to the SM particles through 
a mixing with the CP-odd scalar component from the doublet field. 
There are several ways to avoid appearance of the additional NG boson. 
For example, this NG boson can be absorbed by the additional neutral gauge boson by extending the global symmetry to 
the gauge symmetry via the Higgs mechanism. 
Besides, by introducing explicit breaking terms of the $U(1)$ symmetry, we can avoid such a massless scalar boson. 
In the latter way, if we discuss in the renormalizable theory, 
additional Higgs fields are necessary to construct such an explicit breaking term. 
Of course, we can introduce a higher dimensional term such like $M^{-N}\phi\phi\cdots\phi X$ ($\phi$ is the doublet Higgs field) to break the $U(1)$ symmetry.

\begin{center}
{\bf 5. Examples}
\end{center}

\begin{table}[t]
\begin{center}
{\renewcommand\arraystretch{1.2}
\begin{tabular}{l|l|c|c|c|c|c|c}\hline\hline
Model &Content of Higgs fields&   $\tan\beta$ &$\tan\beta'$ &$\rho_{\text{tree}}$&$|F|^2$& $c_{hWW}$ & $c_{hZZ}$\\\hline
HTM& $\Phi_1=\phi$, $\Phi_2=\Delta$ &$\sqrt{2}v_\Delta/v_\phi$&$2v_\Delta/v_\phi$&$\simeq 1-2v_\Delta^2/v^2$& $c_\beta^2s_\beta^2/c_W^2$& $c_\beta c_\alpha + \sqrt{2}s_\beta s_\alpha$ & $c_{\beta'} c_\alpha + 2s_{\beta'} s_\alpha$ \\\hline
rHTM & $\Phi_1=\phi$, $\Phi_2=\xi$ &$2v_\xi/v_\phi$&-&$ 1+4v_\xi^2/v^2$&$c_\beta^2s_\beta^2/c_W^2$&$c_\beta c_\alpha + 2s_\beta s_\alpha$ & $c_\alpha$\\\hline
GM model & $\Phi_1=\phi$, $\Phi_2=\xi$, $\Phi_3=\Delta$ &$2\sqrt{2}v_T/v_\phi$&$2\sqrt{2}v_T/v_\phi$&1&$s_\beta^2/c_W^2$&$c_\beta c_\alpha +\frac{2\sqrt{6}}{3}s_\beta s_\alpha$ &$c_\beta c_\alpha +\frac{2\sqrt{6}}{3}s_\beta s_\alpha$\\\hline
HSM & $\Phi_1=\phi$, $\Phi_2=\varphi_7$ &$4v_{7}/v_\phi$&$4v_{7}/v_\phi$&1&$s_\beta^2/c_W^2$&$c_\beta c_\alpha +4s_\beta s_\alpha$
&$c_\beta c_\alpha +4s_\beta s_\alpha$ \\\hline\hline
\end{tabular}}
\caption{The coefficients $c_{hVV}$ and $|F|^2$ in various exotic Higgs sectors. 
The mixing angle $\beta$ and $\beta'$ are also given in terms of the VEV's. 
Except for the GM model, $\tan\alpha$ is defined by $R_{2h}/R_{1h}$. 
In the GM model, $R_{1h}$, $R_{2h}$ and $R_{3h}$ are respectively given by $c_\alpha$, $\sqrt{\frac{1}{3}}s_\alpha$ and 
$\sqrt{\frac{2}{3}}s_\alpha$. We use the abbreviations such as $c_X=\cos X$ and $s_X=\sin X$. 
}
\label{models}
\end{center}
\end{table}

We here show that the exotic features discussed in the previous section in several concrete models.  
We define the Higgs fields with the quantum numbers $(T,Y)$ as $\phi$ (1/2,1/2), $\xi$ ($1,0$), $\chi$ ($1,1$) and $\varphi_7$ ($3,2$).
The VEV's for $\phi$, $\xi$, $\Delta$ and $\varphi_7$ are respectively denoted by $v_\phi$, $v_\chi$, $v_\xi$ and $v_7$. 
As examples, we consider the HTM, the real Higgs Triplet Model (rHTM), the GM model and the 
Higgs Septet Model (HSM) whose Higgs fields content is listed in Table~\ref{models}. 
In this table, $\beta$ ($\beta'$) is the mixing angle which separates $G^\pm$ ($G^0$) from 
the physical singly-charged (CP-odd) Higgs bosons. 
In each Higgs sector, 
$\tan\beta$ is calculated by $R_{2G^+}/R_{1G^+}$ ($\sqrt{R_{2G^+}^2+R_{3G^+}^2}/R_{1G^+}$) in the HTM and in the rHTM 
(in the GM model and in the HSM). 
In the HSM, $R_{3G^+}$ corresponds to the singly-charged states of $T_3=-3$ component of $\varphi_7$. 
On the other hand, $\tan\beta'$ is calculated by $R_{2G^0}/R_{1G^0}$ except for in the rHTM because
there is no additional CP-odd scalar state.  

The rho parameter $\rho_{\text{tree}}$, the magnitude of the $H^\pm W^\mp Z$ vertex $|F|^2$ and 
the $hVV$ coupling constant $c_{hVV}$ are listed in Table~\ref{models} in each model. 
In the HTM and the rHTM (In the GM model and in the HSM), 
$\rho_{\text{tree}}$ is not (is) equal to 1, while it is to 1. 
We note that in the GM model, only when the triplet VEV's are aligned as 
\begin{align}
v_T \equiv \frac{v_\Delta}{\sqrt{2}} =v_\xi,  \label{VEV_align}
\end{align} 
then $\rho_{\text{tree}}=1$ is satisfied (see Eq.~(\ref{rho})). 
When $v_\Delta$ ($v_\xi$) is taken to be 3 GeV in the HTM (rHTM), 
the value for $|F|^2$ and the maximum value for $c_{hWW}-1$ and $c_{hZZ}-1$ are obtained as 
$3.9\times 10^{-4}$ ($7.7\times 10^{-4}$), $1.5\times 10^{-4}$ ($8.9\times 10^{-4}$) and $8.9\times 10^{-4}$ (0), respectively in the HTM (rHTM). 
In the GM model and in the HSM, there is no constraint for the VEV's from the rho parameter. 
As an example, when we take $v_T=v_7=30$ GeV, we obtain 
the value for $|F|^2$ and the maximum value for $c_{hVV}-1$ are obtained as 
$0.15$ ($0.31$), $9.4\times 10^{-2}$ ($1.1$), respectively in the GM model (HSM). 

\begin{wrapfigure}{!t}{50mm}
 \begin{center}
  \includegraphics[width=60mm]{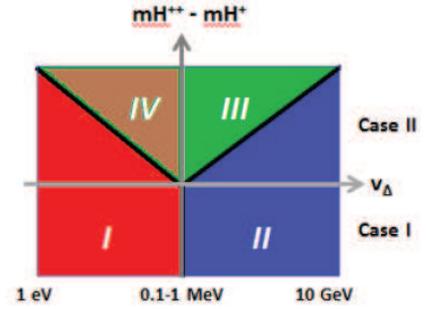}
  \caption{Four regions are schematically shown on the $v_\Delta$-$\Delta m$ plane.}
 \label{fig1}
 \end{center}
\end{wrapfigure}

\section{The Higgs Triplet Model}

In this section, we discuss the HTM in which the complex triplet Higgs field $\Delta$ is added to the SM. 
The most general  
Higgs potential under the $SU(2)_L\times U(1)_Y$ gauge invariance is given by 
\begin{align}
& V = m^2 \phi^\dagger \phi + M^2 {\rm Tr}(\Delta^\dagger \Delta) +
  [ \mu \phi^T i \tau_2 \Delta^\dagger \phi + {\rm h.c.} ] + \lambda_1 (\phi^\dagger \phi)^2\notag\\
&  + \lambda_2 \left[{\rm
                                                  Tr}(\Delta^\dagger
                                                  \Delta) \right]^2
  + \lambda_3 {\rm Tr}\left[(\Delta^\dagger \Delta)^2\right]
  +\lambda_4 (\phi^\dagger\phi) {\rm Tr}(\Delta^\dagger \Delta)
  +\lambda_5 \phi^\dagger \Delta \Delta^\dagger \phi. \notag
 \end{align}  
There are seven physical scalar states; $i.e.$, 
the doubly-charged $H^{\pm\pm}~(=\Delta^{\pm\pm})$, 
the singly-charged $H^\pm$, a CP-odd $A$ as well as two CP-even ($H$ and $h$) scalar states. 
As we already discussed in the previsous sections, the VEV of the triplet field $v_\Delta$ has to be 
much smaller than that of the doublet $v_\phi$ due to the constraint from the electroweak rho parameter. 
In the case of $v_\Delta/v_\phi\ll 1$, $H^{\pm\pm}$, $H^\pm$, $A$ and $H$ ($h$) can be regarded as the triplet-like Higgs bosons 
(the SM-like Higgs boson), and their masses are given by 
\begin{align}
 m_{H^{++}}^2 &= M_\Delta^2- \frac{1}{2}\lambda_5
  v^2,\quad  m_{H^+}^2 = M_\Delta^2 - \frac{1}{4} \lambda_5 v^2,\quad 
 m_A^2 = m_H^2 = M_\Delta^2 ,\notag\\
 m_h^2 &= 2v^2\lambda_1,~\text{with}~M_\Delta^2= \frac{\mu v_\phi^2}{\sqrt{2} v_\Delta}, 
\end{align}
where we neglect the terms proportional to $v_\Delta$. 
Through the new Yukawa interaction $h_{ij}\overline{L_L^{ic}}i\tau_2\Delta L_L^j+\text{h.c.}$
and the $\mu$ term in the potential, the Majorana masses for neutrinos are obtained by 
\begin{align}
(\mathcal{M}_\nu)_{ij}=\sqrt{2}h_{ij}v_\Delta=h_{ij}\frac{\mu v_\phi^2}{M_\Delta^2}.  \label{eq:mn}
\end{align}
It can be seen that when the lepton number violating parameter $\mu$ is taken to be of $\mathcal{O}(0.1-1)$ eV with $h_{ij}=\mathcal{O}(1)$, 
we can take $M_\Delta$ to be $\mathcal{O}(100-1000)$ GeV with satisfying $(\mathcal{M}_\nu)_{ij}=\mathcal{O}(0.1)$ eV 
which is required by the data. 
In such a case, the HTM can be tested at TeV-scale collider experiments. 
There are two ways to test the HTM at collider experiments; namely, (i) the direct way and (ii) the indirect way. 

\begin{center}
{\bf Direct way for testing the HTM}
\end{center}

Discovery of the triplet-like Higgs bosons can be the direct evidence of the HTM. 
In particular, appearance of $H^{\pm\pm}$ is the striking feature of the model so that 
the detection of $H^{\pm\pm}$ is quite important. 
Furthermore, testing the mass spectrum of the triplet-like Higgs boson can be a probe of the HTM, because 
there appear characteristic relationships among the masses as~\cite{AKY} 
\begin{align}
m_{H^{++}}^2-m_{H^{+}}^2 &= m_{H^+}^2-m_A^2~(=-\frac{\lambda_5}{4}v^2) ,~\text{and}~
m_H^2 = m_A^2. \label{mass_relations}
\end{align}
We can see that there are three patterns of the mass spectrum for the triplet-like Higgs bosons. 
In the case with $\lambda_5=0$, all the triplet-like Higgs bosons are degenerate in mass, while
in the case of $\lambda_5>0$ ($\lambda_5<0$), the mass spectrum is 
$m_{A}>m_{H^+}>m_{H^{++}}$~\cite{HTM_mass_diff,Akeroyd_Sugiyama} ($m_{H^{++}}>m_{H^+}>m_A$)~\cite{HTM_mass_diff,AKY}. 

The decay property of $H^{\pm\pm}$ can be drastically different depending on the value of $v_\Delta$ and the mass difference $\Delta m \equiv m_{H^{++}}-m_{H^+}$. 
In the light $H^{\pm\pm}$ case; $i.e., $ $m_{H^{++}}=\mathcal{O}(100)$ GeV, 
the main decay mode of $H^{\pm\pm}$ in the four regions shown in FIG.~\ref{fig1} is expressed as 
\begin{align}
&H^{\pm\pm}\to \ell^\pm\ell^\pm~\text{in Region~I},\quad H^{\pm\pm}\to W^\pm W^\pm~\text{in Region~II},\notag\\
&H^{\pm\pm}\to H^\pm W^{\pm*}\to H/A W^{\pm*}W^{\pm*}\to  b\bar{b}W^{\pm*}W^{\pm*} ( \nu\nu W^{\pm*}W^{\pm*})~\text{in Region~III (in Region~IV)}.
\end{align}
Region~I is the most promising scenario to detect $H^{\pm\pm}$ because of the clear same-sign dilepton signature~\cite{Region1,Han}. 
The structure of the neutrino mass matrix given in Eq.~(\ref{eq:mn}) can be tested by measuring the branching fraction of the $H^{\pm\pm}\to \ell^\pm \ell^\pm$ mode, because its magnitude is determined by $|h_{ij}|^2$. 
However, this scenario has already been excluded by the LHC data if $m_{H^{++}}\lesssim 400$ GeV~\cite{400GeV}. 
In Region~II, $H^{\pm\pm}$ can dominantly decay into the same-sign diboson~\cite{Han,Region2}. 
According to the reference~\cite{Region2}, $H^{\pm\pm}$ with a mass of 180 GeV 
can be tested at 5$\sigma$ level in this scenario with an integrated luminosity of 10 fb$^{-1}$ at 8 TeV. 
Region~III is an important scenario to test the mass relation given in Eq.~(\ref{mass_relations}). 
In this scenario, $H^{\pm\pm}$, $H^\pm$ and $A/H$ may be reconstructed by using the invariant mass as well as 
the transverse mass distributions in the systems of $\ell^\pm\ell^\pm bb E_{T}^{\text{miss}}$, 
$\ell^\pm b\bar{b} E_{T}^{\text{miss}}$ and $b\bar{b}$, respectively  at the LHC~\cite{AKY}. 
Region~IV is a night mare scenario for the detection of $H^{\pm\pm}$, because the decay product of $H^{\pm\pm}$ always include neutrinos.

\begin{center}
{\bf Indirect way for testing the HTM}
\end{center}

Measuring the deviations in coupling constants of the SM-like Higgs boson $h$ from the SM predictions can be an indirect evidence 
of the extended Higgs sector. 
In particular, the Higgs to the diphoton mode $h\to \gamma\gamma$ is one of the most important channels for the SM Higgs boson search 
at the LHC because of the clear signature. 
The decay rate of $h\to \gamma\gamma$ can be modified 
by the loop effect of $H^{\pm\pm}$ and $H^\pm$ as well, which has been calculated in 
the several papers in the HTM~\cite{hgg_HTM,KY}. 

In addition to the $h\to \gamma\gamma$ decay which corresponds to measuring the $h\gamma\gamma$ coupling, 
studying the deviations in $hWW$, $hZZ$ and $hhh$ vertices from the SM predictions are also important. 
These Higgs boson couplings may be accurately measured at future colliders such as the LHC with 3000 fb$^{-1}$ and at the ILC~\cite{Peskin}. 
In Refs.~\cite{AKKY,AKKY_full}, the renormalized Higgs boson couplings $hWW$, $hZZ$ and $hhh$ and also
the decay rate of $h\to \gamma\gamma$ have been calculated at the one-loop level in the HTM. 
It has been found that there are strong correlations among deviations in these Higgs boson couplings. 
For example, if the event number of the $pp\to h\to\gamma\gamma$ channel deviates by $+30\%$ ($-40\%$)
from the SM prediction, deviations in the one-loop corrected $hVV$ and $hhh$ vertices
are predicted about $-0.1\%$ ($-2\%$) and $-10\%$ $(+150\%)$, respectively~\cite{AKKY_full} 
without contradiction with the constraints from the 
vacuum stability~\cite{VS_arhrib} and the perturbative unitarity~\cite{Aoki_Kanemura,VS_arhrib}.

\section{Conclusion}

We have discussed Higgs sectors with exotic representation fields whose isospin are larger than 1/2. 
In such an exotic Higgs sector, there are several characteristic features which do not appear in usual Higgs sectors such as the 
multi-doublet model at the tree level. 
For instance, the electroweak rho parameter can deviate from unity, 
the $H^\pm W^\mp Z$ vertex appears, the $hVV$ vertex can be larger than that in the SM. 
These properties have been seen in concrete models; the HTM, the rHTM, the GM model, and the HSM. 
We also have discussed how to test the HTM at collider experiments, and we have shown two ways; $i.e.,$ 
the direct way and the indirect way.

\bigskip 
{\it Acknowledgments}

I would like to thank Mayumi Aoki, Shinya Kanemura and Mariko Kikuchi for fruitful collaborations. 
I am also grateful to the organizers of the conference HPNP2013 at University of Toyama for their warm
hospitality and giving me the opportunity to have a talk.
This work was supported in part by the National Science Council of R.O.C. under Grant No. NSC-101-2811-M-008-014.


\bigskip 

\begin{thebibliography}{99}

\bibitem{kky}
  S.~Kanemura, M.~Kikuchi and K.~Yagyu,
  arXiv:1301.7303 [hep-ph].

\bibitem{AKKY_full} 
  M.~Aoki, S.~Kanemura, M.~Kikuchi and K.~Yagyu,
  Phys.\ Rev.\ D {\bf 87}, 015012 (2013)
  [arXiv:1211.6029 [hep-ph]].

\bibitem{AKKY} 
  M.~Aoki, S.~Kanemura, M.~Kikuchi and K.~Yagyu,
  Phys.\ Lett.\ B {\bf 714}, 279 (2012)
  [arXiv:1204.1951 [hep-ph]].

\bibitem{KY} 
  S.~Kanemura and K.~Yagyu,
  Phys.\ Rev.\ D {\bf 85}, 115009 (2012)
  [arXiv:1201.6287 [hep-ph]].

\bibitem{AKY} 
  M.~Aoki, S.~Kanemura and K.~Yagyu,
  Phys.\ Rev.\ D {\bf 85}, 055007 (2012)
  [arXiv:1110.4625 [hep-ph]].


\bibitem{Higgs_LHC}
  G.~Aad {\it et al.}  [ATLAS Collaboration],
  Phys.\ Lett.\ B {\bf 716}, 1 (2012)
  [arXiv:1207.7214 [hep-ex]]; 
  S.~Chatrchyan {\it et al.}  [CMS Collaboration],
  Phys.\ Lett.\ B {\bf 716}, 30 (2012)
  [arXiv:1207.7235 [hep-ex]].

\bibitem{GW}
  S.~L.~Glashow and S.~Weinberg,
  Phys.\ Rev.\  D {\bf 15}, 1958 (1977).

\bibitem{IDM}
  R.~Barbieri, L.~J.~Hall and V.~S.~Rychkov,
  Phys.\ Rev.\ D {\bf 74}, 015007 (2006)
  [hep-ph/0603188].


\bibitem{4type}
  V.~D.~Barger, J.~L.~Hewett and R.~J.~N.~Phillips,
  Phys.\ Rev.\ D {\bf 41}, 3421 (1990);
  Y.~Grossman,
  Nucl.\ Phys.\ B {\bf 426}, 355 (1994)
  [hep-ph/9401311].


\bibitem{typeX} 
  M.~Aoki, S.~Kanemura, K.~Tsumura and K.~Yagyu,
  Phys.\ Rev.\ D {\bf 80}, 015017 (2009)
  [arXiv:0902.4665 [hep-ph]];
  V.~Barger, H.~E.~Logan and G.~Shaughnessy,
  Phys.\ Rev.\ D {\bf 79}, 115018 (2009)
  [arXiv:0902.0170 [hep-ph]];
  H.~E.~Logan and D.~MacLennan,
  Phys.\ Rev.\ D {\bf 79}, 115022 (2009)
  [arXiv:0903.2246 [hep-ph]];
  H.~E.~Logan and D.~MacLennan,
  Phys.\ Rev.\ D {\bf 81}, 075016 (2010)
  [arXiv:1002.4916 [hep-ph]]. 

\bibitem{typeIII} 
  T.~P.~Cheng and M.~Sher,
  Phys.\ Rev.\ D {\bf 35}, 3484 (1987).
  D.~Atwood, L.~Reina and A.~Soni,
  Phys.\ Rev.\ D {\bf 55}, 3156 (1997) [hep-ph/9609279]; 
  P.~Ball and R.~Zwicky,
  Phys.\ Rev.\ D {\bf 71}, 014015 (2005)
  [hep-ph/0406232];
  J.~L.~Diaz-Cruz, J.~Hernandez--Sanchez, S.~Moretti, R.~Noriega-Papaqui and A.~Rosado,
  Phys.\ Rev.\ D {\bf 79}, 095025 (2009) [arXiv:0902.4490 [hep-ph]].

\bibitem{pich} 
  A.~Pich and P.~Tuzon,
  Phys.\ Rev.\ D {\bf 80}, 091702 (2009)
  [arXiv:0908.1554 [hep-ph]]. 

\bibitem{recent_THDM} 
  I.~Low, J.~Lykken and G.~Shaughnessy,
  Phys.\ Rev.\ D {\bf 86}, 093012 (2012)
  [arXiv:1207.1093 [hep-ph]];
  H.~S.~Cheon and S.~K.~Kang,
  arXiv:1207.1083 [hep-ph];
  Y.~Bai, V.~Barger, L.~L.~Everett and G.~Shaughnessy,
  arXiv:1210.4922 [hep-ph];
  W.~Altmannshofer, S.~Gori and G.~D.~Kribs,
  Phys.\ Rev.\ D {\bf 86}, 115009 (2012)
  [arXiv:1210.2465 [hep-ph]]; 
  S.~Chang, S.~K.~Kang, J.~-P.~Lee, K.~Y.~Lee, S.~C.~Park and J.~Song,
  arXiv:1210.3439 [hep-ph];
  J.~Chang, K.~Cheung, P.~-Y.~Tseng and T.~-C.~Yuan,
  arXiv:1211.3849 [hep-ph];
  P.~M.~Ferreira, H.~E.~Haber, R.~Santos and J.~P.~Silva,
  arXiv:1211.3131 [hep-ph]; 
  A.~Drozd, B.~Grzadkowski, J.~F.~Gunion and Y.~Jiang,
  arXiv:1211.3580 [hep-ph];
  A.~Celis, V.~Ilisie and A.~Pich,
  arXiv:1302.4022 [hep-ph];
  C.~-W.~Chiang and K.~Yagyu,
  arXiv:1303.0168 [hep-ph];
  B.~inGrinstein and P.~Uttayarat,
  arXiv:1304.0028 [hep-ph]; 
  A.~Barroso, P.~M.~Ferreira, R.~Santos, M.~Sher and J.~P.~Silva,
  arXiv:1304.5225 [hep-ph].

\bibitem{typeII} 
 T.~P.~Cheng and L.~F.~Li,
 Phys.\ Rev.\  D {\bf 22}, 2860 (1980);
 J.~Schechter and J.~W.~F.~Valle,
 Phys.\ Rev.\  D {\bf 22}, 2227 (1980);
  G.~Lazarides, Q.~Shafi and C.~Wetterich,
  Nucl.\ Phys.\  B {\bf 181}, 287 (1981);
  R.~N.~Mohapatra and G.~Senjanovic,
  Phys.\ Rev.\  D {\bf 23}, 165 (1981);
  M.~Magg and C.~Wetterich,
  Phys.\ Lett.\  B {\bf 94}, 61 (1980).



\bibitem{GM} 
  H.~Georgi and M.~Machacek,
  Nucl.\ Phys.\  B {\bf 262}, 463 (1985); 
  M.~S.~Chanowitz and M.~Golden,
  Phys.\ Lett.\  B {\bf 165}, 105 (1985).

\bibitem{GM_pheno}
  J.~F.~Gunion, R.~Vega and J.~Wudka,
  Phys.\ Rev.\  D {\bf 42}, 1673 (1990); 
  R.~Vega and D.~A.~Dicus,
  Nucl.\ Phys.\  B {\bf 329}, 533 (1990);
  J.~F.~Gunion, R.~Vega and J.~Wudka,
  Phys.\ Rev.\  D {\bf 43}, 2322 (1991); 
  H.~E.~Logan and M.~-A.~Roy,
  Phys.\ Rev.\ D {\bf 82}, 115011 (2010)
  [arXiv:1008.4869 [hep-ph]];
  H.~E.~Haber and H.~E.~Logan,
  Phys.\ Rev.\ D {\bf 62}, 015011 (2000)
  [hep-ph/9909335]; 
  C.~-W.~Chiang and K.~Yagyu,
  JHEP {\bf 1301}, 026 (2013)
  [arXiv:1211.2658 [hep-ph]].

\bibitem{Aoki_Kanemura}
  M.~Aoki and S.~Kanemura,
  Phys.\ Rev.\ D {\bf 77}, 095009 (2008)
  [arXiv:0712.4053 [hep-ph]]. 

\bibitem{Kanemura_Sugiyama} 
  S.~Kanemura and H.~Sugiyama,
  Phys.\ Rev.\ D {\bf 86}, 073006 (2012)
  [arXiv:1202.5231 [hep-ph]].


\bibitem{typeII_susy} 
  A.~Rossi,
  Phys.\ Rev.\ D {\bf 66}, 075003 (2002)
  [hep-ph/0207006];
  M.~Hirsch, S.~Kaneko and W.~Porod,
  Phys.\ Rev.\ D {\bf 78}, 093004 (2008)
  [arXiv:0806.3361 [hep-ph]]; 
  T.~Goto, T.~Kubo and Y.~Okada,
  Phys.\ Lett.\ B {\bf 687}, 349 (2010)
  [arXiv:1001.1417 [hep-ph]];
  E.~J.~Chun and P.~Sharma,
  arXiv:1301.1437 [hep-ph].

\bibitem{Quadruplet1} 
  K.~S.~Babu, S.~Nandi and Z.~Tavartkiladze,
  Phys.\ Rev.\ D {\bf 80}, 071702 (2009)
  [arXiv:0905.2710 [hep-ph]]; 
  I.~Picek and B.~Radovcic,
  Phys.\ Lett.\ B {\bf 687}, 338 (2010)
  [arXiv:0911.1374 [hep-ph]].

\bibitem{Quadruplet2} 
  B.~Ren, K.~Tsumura and X.~-G.~He,
  Phys.\ Rev.\ D {\bf 84}, 073004 (2011) [arXiv:1107.5879 [hep-ph]].

\bibitem{Quintuplet} 
  C.~-S.~Chen, C.~-Q.~Geng and L.~-H.~Tsai,
  arXiv:1212.6208 [hep-ph].

\bibitem{Hisano_Tsumura} 
  J.~Hisano and K.~Tsumura,
  arXiv:1301.6455 [hep-ph].

\bibitem{HHG} 
J. F. Gunion, H. E. Haber, G. L. Kane, and S. Dawson,
Front. Phys. {\bf 80}, 1 (2000).


\bibitem{PDG}      
Beringer et al. (Particle Data Group),~PRD~{\bf 86}, 010001 (2012).

\bibitem{Grifols} 
  J.~A.~Grifols and A.~Mendez,
  Phys.\ Rev.\ D {\bf 22}, 1725 (1980).

\bibitem{HWZ_doublet} 
  A.~Mendez and A.~Pomarol,
  Nucl.\ Phys.\  B {\bf 349}, 369 (1991);
  M.~C.~Peyranere, H.~E.~Haber and P.~Irulegui,
  Phys.\ Rev.\  D {\bf 44}, 191 (1991);
  S.~Kanemura,
  Phys.\ Rev.\ D {\bf 61}, 095001 (2000)
  [hep-ph/9710237].


\bibitem{HWZ-LEP}
  D.~K.~Ghosh, R.~M.~Godbole and B.~Mukhopadhyaya,
  Phys.\ Rev.\ D {\bf 55}, 3150 (1997)
  [hep-ph/9605407].

\bibitem{HWZ-Tevatron}
  K.~Cheung and D.~K.~Ghosh,
  JHEP {\bf 0211}, 048 (2002).
 


\bibitem{HWZ-LHC}
  M.~Battaglia, A.~Ferrari, A.~Kiiskinen, T.~Maki,
  [hep-ex/0112015];
  E.~Asakawa and S.~Kanemura,
  Phys.\ Lett.\ B {\bf 626}, 111 (2005)
  [hep-ph/0506310]; 
  E.~Asakawa, S.~Kanemura and J.~Kanzaki,
  Phys.\ Rev.\ D {\bf 75}, 075022 (2007)
  [hep-ph/0612271]; 
  S.~Godfrey and K.~Moats,
  Phys.\ Rev.\ D {\bf 81}, 075026 (2010)
  [arXiv:1003.3033 [hep-ph]].

\bibitem{epem}
  S.~Kanemura,
  Eur.\ Phys.\ J.\ C {\bf 17}, 473 (2000)
  [hep-ph/9911541]; 
  A.~Arhrib, M.~Capdequi Peyranere, W.~Hollik and G.~Moultaka,
  Nucl.\ Phys.\ B {\bf 581}, 34 (2000)
  [Erratum-ibid.\  {\bf 2004}, 400 (2004)]
  [hep-ph/9912527]; 
  S.~Kanemura, S.~Moretti and K.~Odagiri,
  JHEP {\bf 0102}, 011 (2001)
  [hep-ph/0012030]; 
  H.~E.~Logan and S.~-f.~Su,
  Phys.\ Rev.\ D {\bf 66}, 035001 (2002)
  [hep-ph/0203270]; 
  O.~Brein and T.~Hahn,
  Eur.\ Phys.\ J.\ C {\bf 52}, 397 (2007)
  [hep-ph/0610079].

\bibitem{HWZ-ILC}
  S.~Kanemura, K.~Yagyu and K.~Yanase,
  Phys.\ Rev.\ D {\bf 83}, 075018 (2011)
  [arXiv:1103.0493 [hep-ph]].


\bibitem{logan}
H. Logan, 
{\it talk at the ATLAS Canada Workshop, Aug. 21-22 2006, Ottawa, Canada.}; 
  K.~Earl, K.~Hartling, H.~E.~Logan and T.~Pilkington,
  arXiv:1303.1244 [hep-ph].


\bibitem{Han}
  T.~Han, B.~Mukhopadhyaya, Z.~Si and K.~Wang,
  Phys.\ Rev.\ D {\bf 76}, 075013 (2007)
  [arXiv:0706.0441 [hep-ph]];
  P.~Fileviez Perez, T.~Han, G.~-y.~Huang, T.~Li and K.~Wang,
  Phys.\ Rev.\ D {\bf 78}, 015018 (2008)
  [arXiv:0805.3536 [hep-ph]].

\bibitem{Region1} 
  J.~F.~Gunion, C.~Loomis and K.~T.~Pitts, arXiv:hep-ph/9610237; 
M.~Muhlleitner and M.~Spira,
Phys.\ Rev.\ D {\bf 68}, 117701 (2003); 
  M.~Kakizaki, Y.~Ogura and F.~Shima,
  Phys.\ Lett.\ B {\bf 566}, 210 (2003)
  [hep-ph/0304254];
  A.~G.~Akeroyd and M.~Aoki,
  Phys.\ Rev.\ D {\bf 72}, 035011 (2005)
  [hep-ph/0506176];
  M.~Kadastik, M.~Raidal and L.~Rebane,
  Phys.\ Rev.\ D {\bf 77}, 115023 (2008)
  [arXiv:0712.3912 [hep-ph]];
  J.~Garayoa and T.~Schwetz,
  JHEP {\bf 0803}, 009 (2008)
  [arXiv:0712.1453 [hep-ph]]; 
  A.~G.~Akeroyd, M.~Aoki and H.~Sugiyama,
  Phys.\ Rev.\ D {\bf 77}, 075010 (2008)
  [arXiv:0712.4019 [hep-ph]]; 
  A.~G.~Akeroyd and C.~-W.~Chiang,
  Phys.\ Rev.\ D {\bf 80}, 113010 (2009)
  [arXiv:0909.4419 [hep-ph]];
  F.~del Aguila and J.~A.~Aguilar-Saavedra,
  Nucl.\ Phys.\ B {\bf 813}, 22 (2009)
  [arXiv:0808.2468 [hep-ph]]; 
  A.~G.~Akeroyd, C.~W.~Chiang and N.~Gaur, 
  JHEP {\bf 1011}, 005 (2010);
  A.~G.~Akeroyd and C.~-W.~Chiang,
  Phys.\ Rev.\ D {\bf 81}, 115007 (2010)
  [arXiv:1003.3724 [hep-ph]]; 
  E.~J.~Chun and P.~Sharma,
  JHEP {\bf 1208}, 162 (2012)
  [arXiv:1206.6278 [hep-ph]]; 
  H.~Sugiyama, K.~Tsumura and H.~Yokoya,
  Phys.\ Lett.\ B {\bf 717}, 229 (2012)
  [arXiv:1207.0179 [hep-ph]].

\bibitem{400GeV}
  G.~Aad {\it et al.}  [ATLAS Collaboration],
  Phys.\ Rev.\ D {\bf 85}, 032004 (2012)  [arXiv:1201.1091 [hep-ex]];
  S.~Chatrchyan {\it et al.}  [CMS Collaboration],
  Eur.\ Phys.\ J.\ C {\bf 72}, 2189 (2012)
  [arXiv:1207.2666 [hep-ex]].


\bibitem{HTM_mass_diff} 
  S.~Chakrabarti, D.~Choudhury, R.~M.~Godbole and B.~Mukhopadhyaya,
  Phys.\ Lett.\ B {\bf 434}, 347 (1998)
  [hep-ph/9804297];
  E.~J.~Chun, K.~Y.~Lee and S.~C.~Park,
  Phys.\ Lett.\ B {\bf 566}, 142 (2003)
  [hep-ph/0304069]; 
  A.~Melfo, M.~Nemevsek, F.~Nesti, G.~Senjanovic and Y.~Zhang,
  Phys.\ Rev.\ D {\bf 85}, 055018 (2012)
  [arXiv:1108.4416 [hep-ph]].

\bibitem{Akeroyd_Sugiyama} 
  A.~G.~Akeroyd and H.~Sugiyama,
  Phys.\ Rev.\ D {\bf 84}, 035010 (2011)
  [arXiv:1105.2209 [hep-ph]].

\bibitem{Region2} 
  C.~-W.~Chiang, T.~Nomura and K.~Tsumura,
  Phys.\ Rev.\ D {\bf 85}, 095023 (2012)
  [arXiv:1202.2014 [hep-ph]].


\bibitem{hgg_HTM}
  P.~Fileviez Perez, H.~H.~Patel, M.~.J.~Ramsey-Musolf and K.~Wang,
  Phys.\ Rev.\ D {\bf 79}, 055024 (2009)
  [arXiv:0811.3957 [hep-ph]];
  A.~Alves, E.~Ramirez Barreto, A.~G.~Dias, C.~A.~de S.Pires, F.~S.~Queiroz and P.~S.~Rodrigues da Silva,
  Phys.\ Rev.\ D {\bf 84}, 115004 (2011)
  [arXiv:1109.0238 [hep-ph]];
  A.~Arhrib, R.~Benbrik, M.~Chabab, G.~Moultaka and L.~Rahili,
  JHEP {\bf 1204}, 136 (2012)
  [arXiv:1112.5453 [hep-ph]];
   A.~G.~Akeroyd and S.~Moretti,
  Phys.\ Rev.\ D {\bf 86}, 035015 (2012)
  [arXiv:1206.0535 [hep-ph]];
  C.~-W.~Chiang and K.~Yagyu,
  Phys.\ Rev.\ D {\bf 87}, 033003 (2013)
  [arXiv:1207.1065 [hep-ph]];
  E.~J.~Chun, H.~M.~Lee and P.~Sharma,
  JHEP {\bf 1211}, 106 (2012)
  [arXiv:1209.1303 [hep-ph]];
  L.~Wang and X.~-F.~Han,
  Phys.\ Rev.\ D {\bf 87}, 015015 (2013)
  [arXiv:1209.0376 [hep-ph]].

\bibitem{Peskin}
  M.~E.~Peskin,
  arXiv:1207.2516 [hep-ph].





\bibitem{VS_arhrib} 
  A.~Arhrib, R.~Benbrik, M.~Chabab, G.~Moultaka, M.~C.~Peyranere, L.~Rahili and J.~Ramadan,
  Phys.\ Rev.\ D {\bf 84}, 095005 (2011)
  [arXiv:1105.1925 [hep-ph]].




\end{thebibliography}

\end{document}